\documentclass[twocolumn,showpacs,preprintnumbers,amsmath,amssymb]{revtex4}
\usepackage{graphicx}
\begin{document}

\title{Hierarchical-environment-assisted non-Markovian and its effect on thermodynamic properties}

\author{Yong-Wei Li}
\author{Lei Li}
\email{lilei@imu.edu.cn}
\affiliation{School of Physical Science and Technology, Inner Mongolia University, Hohhot 010021, People's Republic of China}

\date{Submitted ****}

\begin{abstract}
We consider a microscopic collision model, i.e., a quantum system interacts with a hierarchical environment consisting of an auxiliary system and a reservoir. We show how the non-Markovian character of the system is influenced by the coupling strength of system-auxiliary system and auxiliary system-reservoir, initial system-environment correlations and the coherence of environment. Then we study the relation between non-Markovianity and thermodynamics properties, by studying the entropy change of system especially that from heat exchanges with memory effects, and we reveal the essence of entropy change between positive and negative values during non-Markovian evolution is due to the contribution of heat flux determined by coherence. And the information flow between the system and environment is always accompanied by energy exchange.
\end{abstract}


\maketitle
\section{Introduction}\label{Sec1}
The study of open quantum systems is of great importance in quantum information and computation recently. Because the dynamics of open quantum systems is greatly affected by its environment, and the environments are often very complex, solving the dynamics of open quantum systems has always been a challenge. The Markovian approximation is important to describe the dynamics of open quantum system either in terms of maps and Kraus operators or in terms of master equations\cite{f1}. One advantage of this approximation is that the dynamics of the system will be a Markovian process and can be described by a standard Markovian master equation.

However, it has been shown that the Markovian approximation fails in many situations\cite{f3,f4,f5,f6}, and the non-Markovian dynamics have been received considerable attention and have been extensively studied recently\cite{f7,f8,f9,f10,f11,qt6,qt7,qt5}. Based on this, several measures of non-Markovianity(NM) have been proposed\cite{f12,f13,f14}. With the help of these measures, one can claim that an evolution is non-Markovian if a nonzero degree of NM is detected. These measures have been applied to many models to investigate their non-Markovian characteristics\cite{f17,f18,f19,f20,f21,f22,f23,r2}. Furthermore, the transition from Markovian to non-Markovian dynamics has also been theoretically and experimentally implemented based on these measures~\cite{f24,f36,ll9,f28,f34,fh,f35}. For example, Brito \textit{et al.} have implemented the transitions from Markovianity to non-Markovianity by preparing different system initial states or dynamically manipulating the subsystem coupling~\cite{f36}. Ma~\textit{et al.} have showed how the non-Markovian character of the system is influenced by the coupling strength between the qubit and cavity and the correlation time of the reservoir, and they have found a phenomenon whereby the qubit Markovian and non-Markovian transition exhibits a anomalous pattern in a parameter space depicted by the coupling strength and the correlation time of the reservoir~\cite{ll9}. In Ref.~\cite{f35}, initial system-environment correlations have been showed to substantially increase the distance between two qubit states evolving to long-time-limit states according to exact non-Markovian dynamics. And in Ref.~\cite{f40}, it have showed that the trace distance between two states of the open system can increase above its initial value when the system and its environment are initially correlated. In particular, Smirne \textit{et al.}~\cite{f41} have provided experimental evidence of the behavior showed in Ref.~\cite{f40}. All of these factors together make it difficult to understand their independent role in the non-Markovian dynamics of open quantum system. However we have not seen any reports about the effect of coherence of environment on the non-Markovian dynamics. Thus an interesting question concerns how the independent role of these factors to influence the system dynamics, specifically system-environment coupling, initial system-environment correlations and the intraenvironment coherence.

As one of the representative models for studying open quantum systems, collision model, also called repeated interaction framework, has been extensively studied during recent decades~\cite{f25,w1,qt1,f44,f26,w12,w13}. A quantum collision model is a microscopic framework to describe the open dynamics of a system interacting with a reservoir assumed to consist of a large collection of smaller constituents (ancillas), and the system is assumed to interact (collide) sequentially with an ancilla at each time step~\cite{f25,f27,qt2}. It offers a bottom-top description of an environment, where one has precise theoretical control of the microscopic aspects that give rise to macroscopic characteristics of the reservoir. The collision model has been applied in non-Markovian dynamics widely~\cite{w11,f46,w9,w10,f43,w4,w5,qt3,qt4,c2,w8,ll10,lil,ycs,jj}. For example, Ciccarello~\textit{et al.} have endowed the reservoir with memory by introducing interancillary collisions between next system-ancilla interactions~\cite{f46}. Bernardes~\textit{et al.} have investigated the Markovian to non-Markovian transitions in collision models by introducing correlations in the state of the environment~\cite{w10}. In Ref.~\cite{f43}, the use of collision model with interenvironment swaps has displayed a signature of strongly non-Markovian dynamics that is highly dependent on the establishment of system-environment correlations. Campbell~\textit{et al.} have also identified the relevant system-environment correlations that lead to a non-Markovian evolution in a collision model~\cite{c2}. Kretschmer~\textit{et al.} have studied the applicability of collisional models for non-Markovian dynamics of open quantum systems, and they have discussed the possibility to embed non-Markovian collision model dynamics into Markovian collision model dynamics in an extended state space~\cite{w4}. Lorenzo~\textit{et al.} have shown that the composite quantum collision models they studied can accommodate some known relevant instances of non-Markovian dynamics~\cite{w8}. In Ref.~\cite{ll10}, a non-Markovian dynamics is established under a structured environment based on collision model. In Ref.~\cite{ycs}, it has studied the effects of different strategies of system-environment interactions and states of the blocks on the non-Markovianities by introducing a block (a number of environment particles) as the unit of the environment instead of a single particle. Ref.~\cite{jj} has found that the information is scrambled if the memory and environmental particles are alternatively squeezed along two directions which are perpendicular to each other.

Recently, the relation between non-Markovianity and thermodynamics in open quantum system has attracted considerable attention. In Ref.~\cite{w7}, the heat flux has exhibited a nonexponential time behavior in the case of non-Markovian dynamics of the subsystem. In Ref.~\cite{w2}, the heat flux changes between positive to negative values for a non-Markovian evolution of the subsystem, which leads to a violation of open-system formulation of Landauer's principle for the heat and entropy fluxes. A similar result has also been obtained in Refs.~\cite{f42,e6} that the Landauer's principle is violated in non-Markovian dynamics. Raja~\textit{et al.} have investigated how memory effects influence the ability to perform work on the driven qubit, and they have showed that the average work performed on the qubit can be used as a diagnostic tool to detect the presence or absence of memory effects~\cite{f34}. Pezzutto~\textit{et al.} have addressed the effects that non-Markovianity of the open-system dynamics of the work medium can have on the efficiency of the thermal machine~\cite{e7}. Katz~\textit{et al.} have studied the performance characteristics of a heat rectifier and a heat pump in a non-Markovian framework~\cite{r1}. Ref.~\cite{e8} has studied the effects of environmental temperature on the non-Markovianity of an open quantum system by virtue of collision models.

In this paper, we consider a two-level system coupled to a structured environment consisting of a auxiliary system and a reservoir, and the reservoir is of a large collection of initially uncorrelated systems which we call ancillas (see figure 1). Based on this structured environment model, there can be different approaches to influence the non-Markovian character of the system, such as the coupling strength of system-auxiliary system and auxiliary system-reservoir, initial system-environment correlations and the coherence of environment. And the non-Markovianity can influence the thermodynamic properties of system. For example, the entropy change of system is between positive and negative values during non-Markovian evolution, and the essence of this is due to the contribution of heat flux determined by coherence. And the information flow between the system and environment is always accompanied by energy exchange.

\section{Model and solution}\label{Sec2}
We consider a qubit (system $S$) couples to a hierarchical environment, which contains a
auxiliary qubit $A_{Q}$ and a collection of $N$ identical noninteracting ancillas
(qubits) $\{\mathcal{R}_{1},\mathcal{R}_{2},...,\mathcal{R}_{N}\}$ that consists a reservoir $\mathcal{R}$, and this reservoir is in
the product state $\eta_{tot}=\otimes^{N}_{j=1}\eta_{j}$. In this way, the auxiliary qubit $A_{Q}$ and the
reservoir hierarchically constitute the whole big reservoir $E$, which is called the environment of system $S$. And the general scheme is illustrated in Fig. 1. The Hamiltonians of system and a generic environment particle $E_{j}$ including the auxiliary qubit and ancillas are
\begin{equation}
\hat{H}_{S(E)}=\omega_{S(E)}\hat{\sigma}_{z}/2,
\end{equation}
where $\hat{\sigma}_{z}$ is the the Pauli matrices and we set $\hbar=1$ throughout this paper.

\begin{figure}[h]
\includegraphics[scale=0.56]{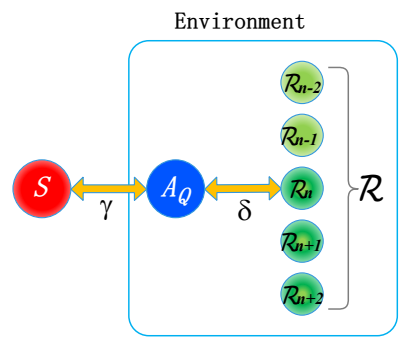}
\centering
\caption{(Color online) Sketch of the protocol of system $S$ plus a hierarchical environment. $S$ interacts with the environment: After $A_{Q}$ interacts with $\mathcal{R}_{n}$ (the $n$th ancilla of reservoir $\mathcal{R}$), it collides with $S$ and is then directed to $\mathcal{R}_{n+1}$.}
\label{Fig1}
\end{figure}

The evolution of system $S$ and its interaction with the environment are proceeded as follows.
$S$ interacts with the environment first: Specifically $S$ and $A_{Q}$ interact and then
$A_{Q}$ collides with the individual ancilla of the reservoir. As the assumption of a big reservoir $\mathcal{R}$
that $A_{Q}$ never interacts twice with the same ancilla, i.e., at each collision the state of the ancilla is refreshed.
And this process is implemented through the unitary operator
\begin{equation}
\hat{\mathcal{U}}_{SE}=\hat{V}_{A_{Q},\mathcal{R}_{j}}\hat{U}_{S,A_{Q}},
\end{equation}
where $\hat{U}_{S,A_{Q}}=e^{-i\hat{H}^{int}_{S,A_{Q}}\tau}$, $\hat{V}_{A_{Q},\mathcal{R}_{j}}=e^{-i\hat{H}^{int}_{A_{Q},\mathcal{R}_{j}}\tau}$.
Here $\hat{H}^{int}_{S,A_{Q}}$ and $\hat{H}^{int}_{A_{Q},\mathcal{R}_{j}}$ are the interaction between `$S-A_{Q}$', `$A_{Q}-\mathcal{R}_{j}$' respectively, and $\tau$ is the interaction time.

In our model, we consider a coherent interaction between the bipartite systems including `$S-{A}_{Q}$' and `$A_{Q}-\mathcal{R}_{j}$',
i.e., a mechanism that can be described by a Hamiltonian model of some form, specifically in this paper we suppose that the
interaction Hamiltonian is
\begin{equation}
\begin{split}
\hat{H}^{int}_{S,A_{Q}(A_{Q},\mathcal{R}_{j})}&=g_{1(2)}(\hat{\sigma}^{S(A_{Q})}_{x}\hat{\sigma}^{A_{Q}(\mathcal{R}_{j})}_{x}+\hat{\sigma}^{S(A_{Q})}_{y}\hat{\sigma}^{A_{Q}(\mathcal{R}_{j})}_{y} \\
&+\hat{\sigma}^{S(A_{Q})}_{z}\hat{\sigma}^{A_{Q}(\mathcal{R}_{j})}_{z}),
\end{split}
\end{equation}
where $\hat{\sigma}^{S}_{i}$, $\hat{\sigma}^{A_{Q}}_{i}$ and $\hat{\sigma}^{\mathcal{R}_{j}}_{i}$ $(i=x,y,z)$ are the Pauli matrices, and $g_{1(2)}$ is a coupling constant. And we use the result~\cite{f25}
\begin{equation}
e^{i\frac{\phi}{2}(\hat{\sigma}_{x}\otimes\hat{\sigma}_{x}+\hat{\sigma}_{y}\otimes\hat{\sigma}_{y}+\hat{\sigma}_{z}\otimes\hat{\sigma}_{z})}=e^{-i\frac{\phi}{2}}(\cos\phi\hat{\mathbb{I}}+i\sin\phi\hat{S}^{sw}),
\end{equation}
where $\hat{\mathbb{I}}$ is the identity operator, and $\hat{S}^{sw}$ is the two-particle swap operator, i.e., it is the unitary operation whose action is $|\psi_{1}\rangle\otimes|\psi_{2}\rangle\rightarrow|\psi_{2}\rangle\otimes|\psi_{1}\rangle$ for all $|\psi_{1}\rangle,|\psi_{2}\rangle\in\mathbb{C}^{2}$. We can now write the unitary time-evolution operator $\hat{U}_{S,A_{Q}}$ in Eq. (2) as
\begin{equation}
\hat{U}_{S,A_{Q}}(\gamma)=(\cos\gamma)\hat{\mathbb{I}}_{S,A_{Q}}+i(\sin\gamma)\hat{S}_{S,A_{Q}}^{sw},
\end{equation}
where $\gamma=2g_{1}\tau$ is a dimensionless interaction strength. And when $\gamma=0$ Eq. (5) is reduced into an identity operator and indicates that there is no interaction between $S$ and $A_{Q}$; and when $\gamma=\pi/2$ Eq. (5) is reduced into a fully swap operator and represents a complete exchange of quantum state information between $S$ and $A_{Q}$. Thus in the range of $\gamma\in[0,\pi/2]$, the larger the $\gamma$, the stronger the coupling. And in the ordered basis $\{|00\rangle, |01\rangle, |10\rangle, |11\rangle\}$, $\hat{S}_{S,A_{Q}}^{sw}$ in Eq. (5) reads~\cite{ll2}
\begin{equation}
\hat{S}_{S,A_{Q}}^{sw}=
\begin{pmatrix}
1& \hspace{0.25cm}0 & \hspace{0.25cm}0 & \hspace{0.25cm}0\\
\*0 &\hspace{0.25cm}0 & \hspace{0.25cm}1& \hspace{0.25cm}0\\
\*0 &\hspace{0.25cm}1 & \hspace{0.25cm}0& \hspace{0.25cm}0\\
0& \hspace{0.25cm}0 & \hspace{0.25cm}0 & \hspace{0.25cm}1%
\end{pmatrix}%
\label{initial}.
\end{equation}
Similarly $\hat{V}_{A_{Q},\mathcal{R}_{j}}$ in Eq. (2) can be written as
\begin{equation}
\hat{V}_{A_{Q},\mathcal{R}_{j}}(\delta)=(\cos\delta)\hat{\mathbb{I}}_{A_{Q},\mathcal{R}_{j}}+i(\sin\delta)\hat{S}_{A_{Q},\mathcal{R}_{j}}^{sw},
\end{equation}
with $\delta\neq\gamma$, in general, and the analog of the operations introduced above applies to $\hat{\mathbb{I}}_{A_{Q},\mathcal{R}_{j}}$ and $\hat{S}^{sw}_{A_{Q},\mathcal{R}_{j}}$ (swap gate between $A_{Q}$ and $\mathcal{R}_{j}$). As mentioned above the dynamics of system $S$ consists of sequential system-environment interaction and each step is treated in the following process: First $S$ and $A_{Q}$ interact and then subsequently $A_{Q}$ collides with $\mathcal{R}_{j}$ (one of the ancillas in $\mathcal{R}$). Thus the system is brought from step $n$ to step $n+1$ through the process
\begin{equation}
\rho^{S,{A}_{Q}}_{n}\otimes\eta_{n+1}\rightarrow \rho^{SE}_{n+1}=\hat{\mathcal{U}}_{SE}(\rho^{S,{A}_{Q}}_{n}\otimes\eta_{n+1})\hat{\mathcal{U}}_{SE}^{\dagger},
\end{equation}
where $\rho^{S,A_{Q}}_{n}$ is the state of `$S-A_{Q}$' after the $n$th interaction. Hence after the $(n+1)$th interaction, we can obtain
the reduced system state, $\rho^{S,A_{Q}}_{n+1}=\mathrm{Tr}_{\mathcal{R}}[\rho^{SE}_{n+1}]$ (the state of `$S-A_{Q}$'),
$\rho^{S}_{n+1}=\mathrm{Tr}_{A_{Q}}[\rho^{S,A_{Q}}_{n+1}]$ (the state of $S$) and $\rho^{A_{Q}}_{n+1}=\mathrm{Tr}_{S}[\rho^{S,A_{Q}}_{n+1}]$ (the state of $A_{Q}$), where $\mathrm{Tr}_{x}[\cdots]$ means the trace of $x$ degree of freedom.

\section{Non-Markovian dynamics of system $S$}\label{Sec3}
The trace distance between two quantum states is one of the most important measures of distinguishability of quantum states~\cite{ll2}, which is given by
\begin{equation}
\mathcal{D}(\rho_{1},\rho_{2})=\frac{1}{2}\mathrm{Tr}|\rho_{1}-\rho_{2}|,
\end{equation}
where $|A|=\sqrt{A^{\dagger}A}$ for any operator $A$. It is obvious that for any pair of states $\rho_{1}$ and $\rho_{2}$ the trace distance satisfies the inequality $0\leqslant\mathcal{D}(\rho_{1},\rho_{2})\leqslant1$. For the time evolution of a quantum state described by a trace-preserving completely positive map, the trace distance is always less than or equal to the initial value~\cite{ll3}; that is,
\begin{equation}
\mathcal{D}(\rho_{1}(t),\rho_{2}(t))\leqslant\mathcal{D}(\rho_{1}(0),\rho_{2}(0)).
\end{equation}
In particular, for a Markovian evolution it can always be represented by a dynamical semigroup of completely positive and trace-preserving maps~\cite{ll4}, and we obtain the inequality
\begin{equation}
\mathcal{D}(\rho_{1}(t+\tau),\rho_{2}(t+\tau))\leqslant\mathcal{D}(\rho_{1}(t),\rho_{2}(t)),
\end{equation}
for any positive $\tau$, which indicates that the trace distance decreases monotonically with time. The decrease of trace distance corresponds to the reduction of distinguish ability between the two states, and this could be interpreted as an outflow of information from the system to the environment.
In contrast to this, if the time derivative of the trace distance becomes positive in some time intervals, the time evolution is non-Markovian~\cite{f12,f17}. Furthermore, if the trace distance exceeds the initial value, the time evolution cannot be described by a trace-preserving completely positive map. Based on this, a measure of non-Markovianity (NM) can be defined by~\cite{f12}
\begin{equation}
\mathcal{N}=\mathop{max}\limits_{\rho_{1}(0),\rho_{2}(0)}\int_{\sigma>0}\mathrm{d}t\ \sigma(t,\rho_{1}(0),\rho_{2}(0)),
\end{equation}
where $\sigma(t,\rho_{1}(0),\rho_{2}(0))=\frac{\mathrm{d}}{\mathrm{d}t}\mathcal{D}(\rho_{1}(t),\rho_{2}(t))$. Conceptually, $\mathcal{N}$ accounts for all regions where the distance between two arbitrary input states increases, thus witnessing a backflow of information from the environment to system. And in this case, an evolution is non-Markovian if and only if $\mathcal{N}>0$.

As the evolution in our model proceeds in discrete steps, we will employ the discretized version of Eq. (12), which is obtained as~\cite{f17,ll8}
\begin{equation}
\mathcal{N}=max\sum_{n\in\sigma^{+}}[\mathcal{D}(\rho_{1,n+1},\rho_{2,n+1})-\mathcal{D}(\rho_{1,n},\rho_{2,n})],
\end{equation}
with $\sigma^{+}=\bigcup_{n}(n,n+1)$ is the union of all the interaction steps $(n,n+1)$ within which $\mathcal{D}(\rho_{1,n+1},\rho_{2,n+1})-\mathcal{D}(\rho_{1,n},\rho_{2,n})>0$, and $\{\rho_{1,n+1},\rho_{2,n+1}\}$ a pair of state of system obtained starting from the corresponding pair of orthogonal state $\{|\psi_{+}\rangle,|\psi_{-}\rangle\}$ after $n+1$ steps of our protocol,
\begin{equation}
\begin{split}
|\psi_{+}\rangle=\cos\frac{\theta}{2}|0\rangle+e^{i\varphi}\sin\frac{\theta}{2}|1\rangle,\\
|\psi_{-}\rangle=\sin\frac{\theta}{2}|0\rangle-e^{i\varphi}\cos\frac{\theta}{2}|1\rangle,
\end{split}
\end{equation}
where $\theta\in[0,\frac{\pi}{2}]$ and $\varphi\in[0,2\pi]$. The maximization in Eq. (13) performed over all possible values of $\theta$ and $\varphi$, i.e., all possible orthogonal pairs of initial system states. In the following we study how the system dynamics can be affected by different ways, including the coupling strength between the bipartite systems (`$S-{A}_{Q}$' and `$A_{Q}-\mathcal{R}$'), coherence of the environment and initial system-environment correlation. Thus, we consider the initial state of each ancilla of reservoir $\mathcal{R}$ as
\begin{equation}
\rho_{coh}=p|\psi\rangle\langle\psi|+(1-p)\rho_{\beta},
\end{equation}
where $p\in[0,1]$, $|\psi\rangle=\frac{1}{\sqrt{Z}}(e^{-\frac{1}{4}\omega_{E}\beta}|0\rangle+e^{i\phi_{1}+\frac{1}{4}\omega_{E}\beta}|1\rangle)$ with a relative phase $\phi_{1}$, and $\rho_{\beta}$ is the thermal state assumed to be of canonical equilibrium form, i.e., $\rho_{\beta}=\frac{1}{Z}e^{-\beta\hat{H}_{E}}$. Here $\beta=1/T$ and $Z=\textmd{Tr}[e^{-\beta\hat{H}_{E}}]$ are the inverse temperature and the partition function respectively. Note that the diagonal elements of states $\rho_{coh}$ and $\rho_{\beta}$ are identical, and compared with the thermal state, the off-diagonal elements of state $\rho_{coh}$ are nonzero if $p\neq0$. Therefore, Eq. (15) can also be written as
\begin{equation}
\rho_{coh}=\rho_{\beta}+p\rho_{non},
\end{equation}
where $\rho_{non}$ is the non-diagonal part of state $|\psi\rangle\langle\psi|$, i.e., the off-diagonal elements of $\rho_{non}$ are the same as that of state $|\psi\rangle\langle\psi|$ and the diagonal elements are zero.

\subsection{Effect of the coupling strength on NM}
In this section we suppose that the environment is in thermal state, i.e., all environment particles including $A_{Q}$ and each ancilla are in the state $\rho_{\beta}$ with $T=\omega_{E}=1$. We numerically calculate the degree of NM for different $\gamma$ and $\delta$ which is presented in Fig. 2. We can see that the whole diagram is divided into two regions, where the green stars represent the degree of NM being equal to zero (Markovian region) and the red dots represent the degree of NM being larger than zero (non-Markovian region). It shows that the non-Markovian dynamics of the system is determined by a delicate balance between the two parameters $\gamma$ and $\delta$. Specifically the system dynamics is Markovian for small $\gamma$ and larger $\delta$, and the non-Markovian region increases with the increase of $\gamma$. Physically this can be understood as following. When the interaction between $S$ and $A_{Q}$ is small (small $\gamma$) and with a relatively large interaction between $A_{Q}$ and $\mathcal{R}_{j}$ (larger $\delta$), the information obtained by $A_{Q}$ from $S$ is less and all of which flow into the reservoir $\mathcal{R}$, which forms Markovian dynamics of the system. In other words, the system is losing information at a slower rate than that of the evolution of environment, thus the backflow of information cannot happen now. However with the increase of $\gamma$, more and more information flows from $S$ into $A_{Q}$ which leads to only part of the information flows into the reservoir and the rest is reserved and flows back to $S$, and in this case the non-Markovian dynamics of system is formed.
\begin{figure}[h]
\includegraphics[scale=0.52]{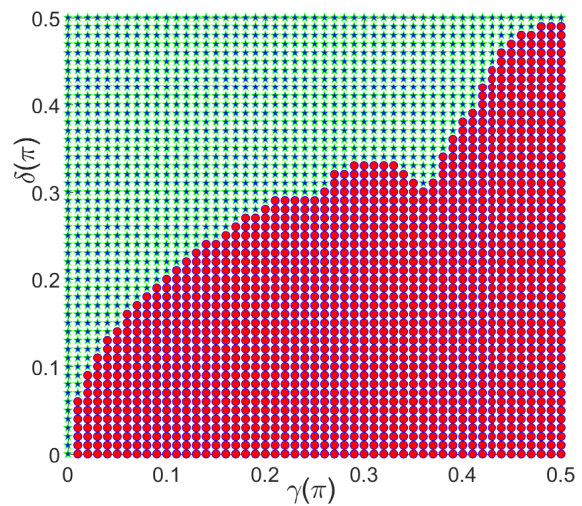}
\centering
\caption{(Color online) The transition from Markovian to non-Markovian dynamics induced by manipulating coupling strength $\gamma$ and $\delta$. The $\mathcal{N}$ in Eq. (13) is performed over all possible $\theta$ and $\varphi$ of initial state (14). And the green stars represent $\mathcal{N}$ being equal to zero (Markovian region) and the red dots represent $\mathcal{N}$ being larger than zero (non-Markovian region).}
\label{Fig2}
\end{figure}

\subsection{Effect of coherence of environment on NM}
In this section we consider the case of environment with coherence, i.e., $A_{Q}$ and each ancilla are in state (16) with a relative phase $\phi_{2}$ and $\phi_{1}$ respectively. Thus the phase difference between reservoir $\mathcal{R}$ and $A_{Q}$ is $\phi=\phi_{1}-\phi_{2}$. According to the degree of NM being equal to zero (Markovian process) or larger than zero (non-Markovian process), we plot the diagram of system dynamics in Fig. 3 with $\varphi$ and $\theta$ in Eq. (14), for different parameter $p$ with fixed phase difference $\phi=0$ (Fig. 3a), and different phase difference $\phi$ with fixed parameter $p$ ($p=0.4$, Fig. 3b), and we set $\gamma=\frac{\pi}{14}$, $\delta=\frac{\pi}{6}$ and $T=\omega_{E}=1$ which is the Markovian region of coupling presented in Fig. 2. From Eq. (16) we know that the coherence of environment is increased with the increase of parameter $p$. From numerical calculations we find that the system dynamics is Markovian for $p\in[0,0.4]$, and the non-Markovian region resulting from environment-coherence is increased with the increase of $p$ in the region $p\in[0.5,1]$ (Fig. 3(a)). This reveals a transition from Markovian to non-Markovian dynamics by coherence of environment, and the larger degree of coherence of environment the easier of non-Markovian dynamics to be obtained. Besides parameter $p$ phase difference $\phi$ is also one of the influence factor of coherence of environment. Thus in Fig. 3(b) we plot the system dynamics with $\varphi$ and $\theta$, for different phase difference $\phi=\{0,\pi/4,\pi/2,\pi,5\pi/4,3\pi/2\}$ with fixed parameter $p=0.4$ (Markovian dynamics presented in Fig. 3(a)). In this case, it can be seen that the system dynamics can also be changed from Markovian to non-Markovian and the non-Markovian region resulting from environment-coherence is different for different $\phi$. From the discussion above, the system dynamics can be changed by means of coherence of environment.
\begin{figure}[h]
\includegraphics[scale=0.52]{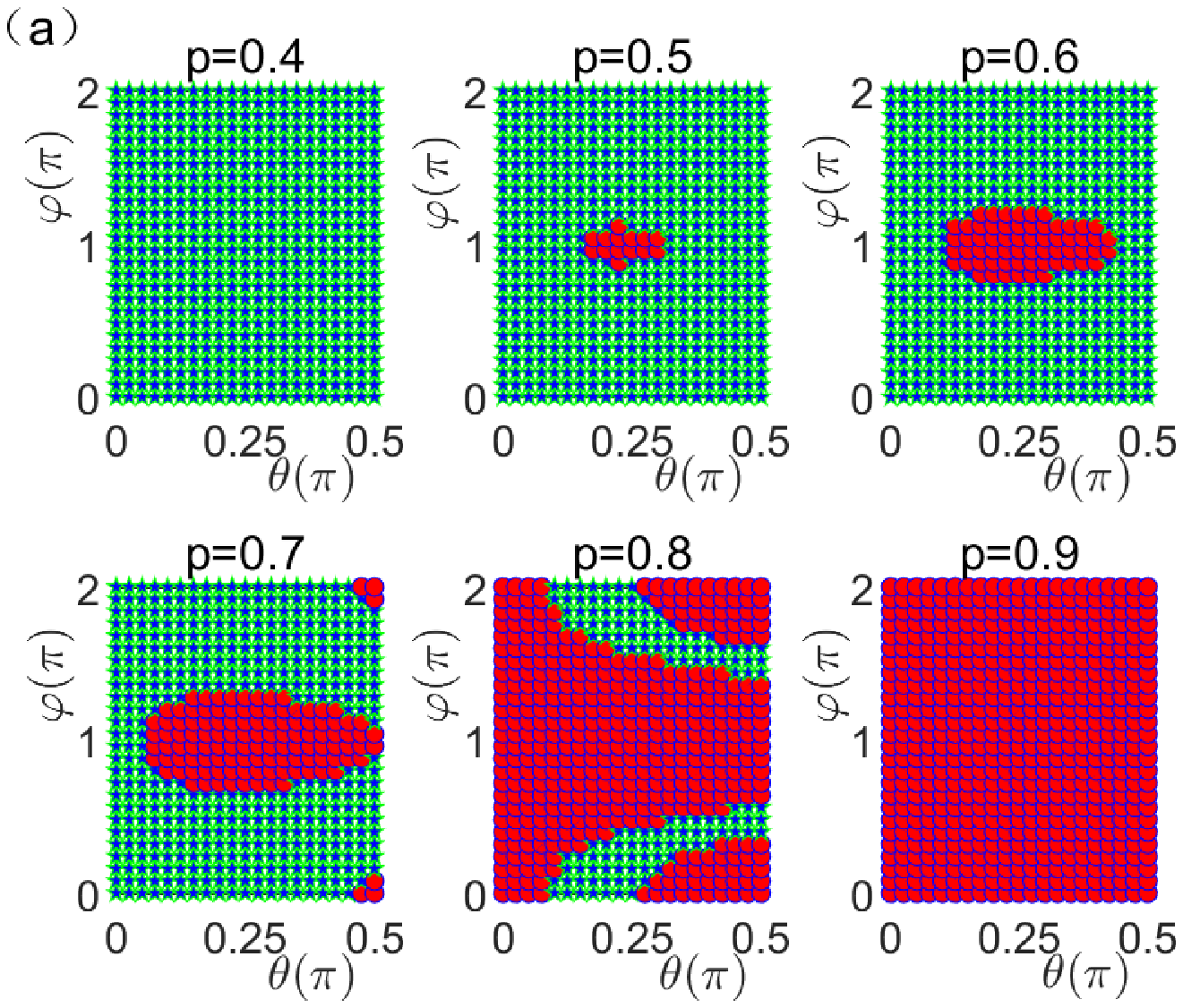}
\includegraphics[scale=0.52]{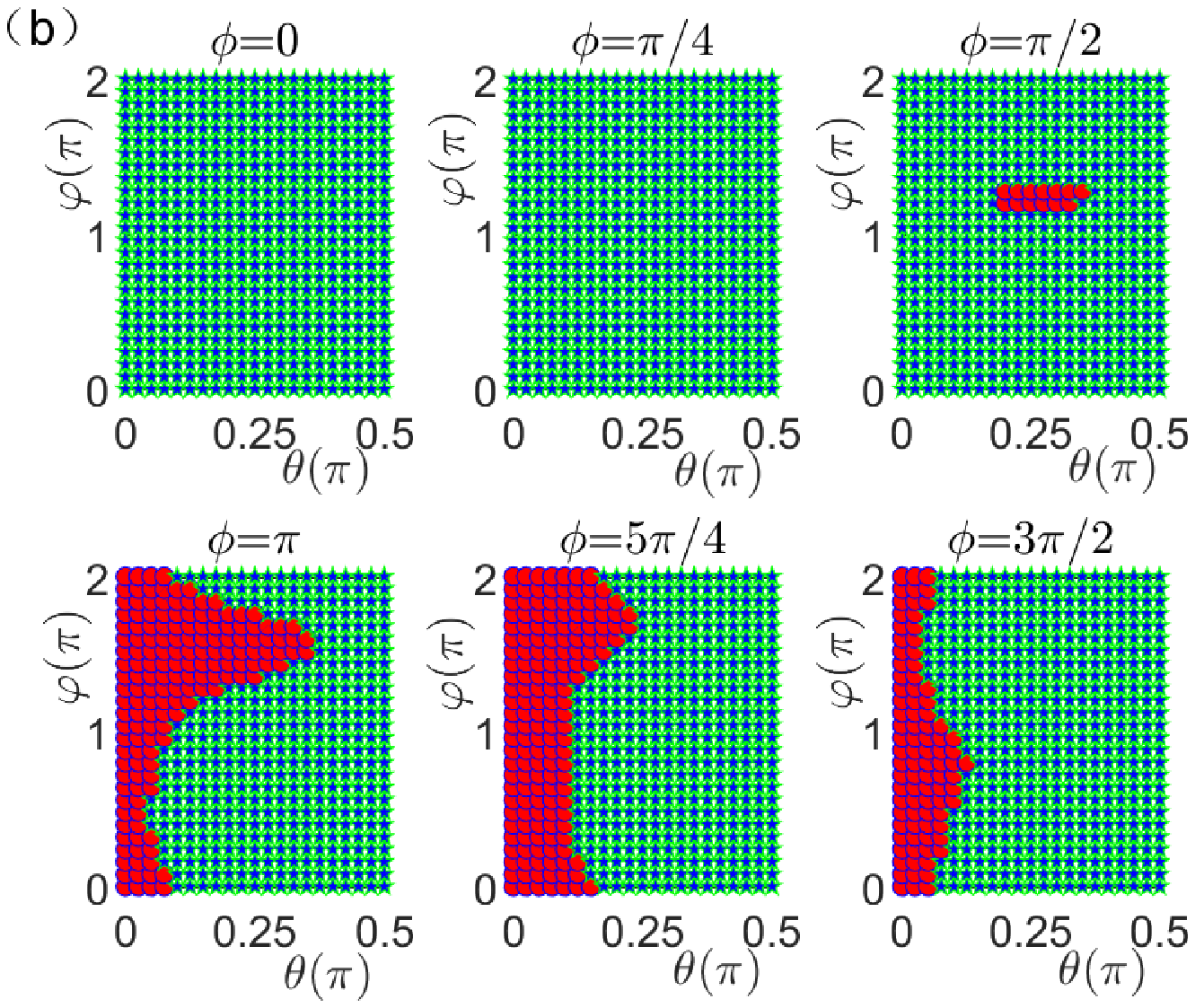}
\centering
\caption{(Color online) The transition from Markovian to non-Markovian dynamics induced by the coherent of environment with fixed phase difference between reservoir $\mathcal{R}$ and $A_{Q}$, $\phi=0$ ($a$), and different phase difference $\phi=\{0,\pi/4,\pi/2,\pi,5\pi/4,3\pi/2\}$ with fixed parameter $p$, $p=0.4$ ($b$).
For both plots the representation of green stars and red dots are the same with Fig. 2, and $\gamma=\frac{\pi}{14}$ and $\delta=\frac{\pi}{6}$ which is the Markovian region of coupling presented in Fig. 2.}
\label{Fig3}
\end{figure}

\subsection{Effect of initial system-environment correlations on NM}
In order to illustrate the effect of initial system-environment correlations on NM, we consider a group of two initial states
\begin{equation}
\rho_{S,A_{Q}}^{1}(0)=|\psi\rangle\langle\psi|=
\begin{pmatrix}
0& \hspace{0.25cm}0 & \hspace{0.25cm}0 & \hspace{0.25cm}0\\
\*0 &\hspace{0.25cm}|\xi|^{2} & \hspace{0.25cm}\xi\sqrt{1-\xi^{2}}& \hspace{0.25cm}0\\
\*0 &\hspace{0.25cm}\xi\sqrt{1-\xi^{2}} & \hspace{0.25cm}1-|\xi|^{2}& \hspace{0.25cm}0\\
0& \hspace{0.25cm}0 & \hspace{0.25cm}0 & \hspace{0.25cm}0%
\end{pmatrix}
\end{equation}
\begin{equation}
\begin{aligned}
\rho_{S,A_{Q}}^{2}(0)&=|\xi|^{2}|0 \rangle\langle 0|\otimes|1 \rangle\langle 1|+(1-|\xi|^{2})|1 \rangle\langle 1|\otimes|0 \rangle\langle 0|\\
&=
\begin{pmatrix}
0& \hspace{0.25cm}0 & \hspace{0.25cm}0 & \hspace{0.25cm}0\\
\*0 &\hspace{0.25cm}|\xi|^{2} & \hspace{0.25cm}0& \hspace{0.25cm}0\\
\*0 &\hspace{0.25cm}0 & \hspace{0.25cm}1-|\xi|^{2}& \hspace{0.25cm}0\\
0& \hspace{0.25cm}0 & \hspace{0.25cm}0 & \hspace{0.25cm}0%
\end{pmatrix}\\
\end{aligned}
\end{equation}
where $|\psi\rangle=|\xi||01\rangle+\sqrt{1-|\xi|^{2}}|10\rangle$. It is worth noting that the reduced density matrixs of these two initial states are identical, and $\rho_{S,A_{Q}}^{1}(0)$ has quantum correlations between $S$ and $A_{Q}$ initially, $\rho_{S,A_{Q}}^{2}(0)$ has classical correlation. And in the case of initial correlation between $S$ and $A_{Q}$, the non-Markovian dynamics of system can be witnessed by the non-monotonicity in evolution of trace distance between two states of system
\begin{equation}
\mathcal{D}(\rho_{S}(n+1),\tilde{\rho}_{S}(n+1)).
\end{equation}
Here $\rho_{S}(n+1)$ and $\tilde{\rho}_{S}(n+1)$ are a pair of states of system after the $(n+1)$th step, which corresponds to the pair of initial states of the composite system `$S-A_{Q}$', $\{\rho_{S,A_{Q}}(0), \tilde{\rho}_{S,A_{Q}}(0)\}$, and
$\tilde{\rho}_{S,A_{Q}}(0)=\mathrm{Tr}_{A_{Q}}(\rho_{S,A_{Q}}(0))\otimes\mathrm{Tr}_{S}(\rho_{S,A_{Q}}(0))$. And in this section we suppose that each ancilla in $\mathcal{R}$ are initially in the thermal state $\rho_{\beta}$ with $T=\omega_{E}=1$.

\begin{figure}[h]
\includegraphics[scale=0.56]{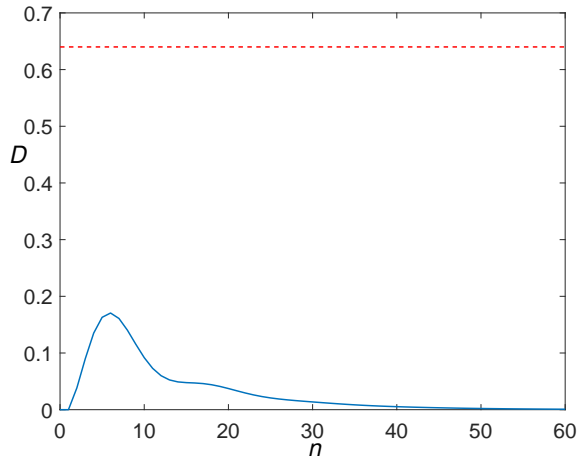}
\centering
\caption{(Color online)
Trace distance Eq. (19) (solid blue line) against the number of collisions $n$, a non-Markovian process induced by initial quantum correlation between system $S$ and
environment in the Markovian coupling region of Fig. 2 with $\{\gamma=\frac{\pi}{14}$, $\delta=\frac{\pi}{6}$\}, and the other parameters are the same with Fig. 2. And the bound of Eq. (20) is indicated by the upper red-dashed line.}
\label{Fig4}
\end{figure}

In Fig. 4 we plot the trace distance Eq. (19) against the number of collisions $n$ for initial state (17) (initial quantum correlations between $S$ and $A_{Q}$) and thermal state of the reservoir, $\gamma=\frac{\pi}{14}$ and $\delta=\frac{\pi}{6}$ which correspond to a Markovian region of coupling presented in Fig. 2, and we set $\xi=0.855$ in Eq. (17) to let the reduced state of $A_{Q}$ is the same with the state of each ancilla. It shows that the trace distance increases from zero to a maximum and then decreases until to zero, which implies that a non-Markovian dynamics of system and the trace distance here exceeds the initial value. Laine $\emph{et al.}$ have pointed out that the trace distance between two states of the open system can increase above its initial value when system and its environment are initially correlated~\cite{f40}. And in our case it can be written as
\begin{equation}
\begin{aligned}
&\mathcal{D}[\rho_{S}(n), \tilde{\rho}_{S}(n)]\leqslant\\
&\mathcal{D}[\rho_{S,A_{Q}}^{1}(0),\mathrm{Tr}_{A_{Q}}(\rho_{S,A_{Q}}^{1}(0))\otimes\mathrm{Tr}_{S}(\rho_{S,A_{Q}}^{1}(0))],
\end{aligned}
\end{equation}
where $\rho_{S}(n)$ and $\tilde{\rho}_{S}(n)$ are the reduced state of system after the $n$th interaction corresponding to the initial state $\rho_{S,A_{Q}}^{1}(0)$ and $\mathrm{Tr}_{A_{Q}}(\rho_{S,A_{Q}}^{1}(0))\otimes\mathrm{Tr}_{S}(\rho_{S,A_{Q}}^{1}(0))$, respectively. This inequality shows how far from each other two initially indistinguishable reduced states can evolve when only one of the two initial states is correlated. And physically this can be understood as following: The maximal amount of information the open system can gain from the environment is the amount of information flowed out earlier from the system since the initial time, plus the information which is initially outside the open system. Thus the increase of the trace distance is bounded from above by the correlations in the initial state. We calculate the bound of Eq. (20) and which is showed by red line in Fig. 4, and the inequality Eq. (20) is well satisfied. We notice that the maximum value of the trace distance at a certain $n$ in Fig. 4 is much smaller than the bound of Eq. (20), i.e., the bound of Eq. (20) is actually loose. This means that only less of the information in the composite system initially transfers to the reduced system during the evolution, and which is due to the Markovian reservoir $\mathcal{R}$. Moreover, Smirne $\emph{et al.}$ have provided experimental evidence that if the environmental state is fixed, the trace distance between two states of an open quantum system can increase over its initial value only in the presence of initial correlations~\cite{f41}.

\begin{figure}[h]
\includegraphics[scale=0.56]{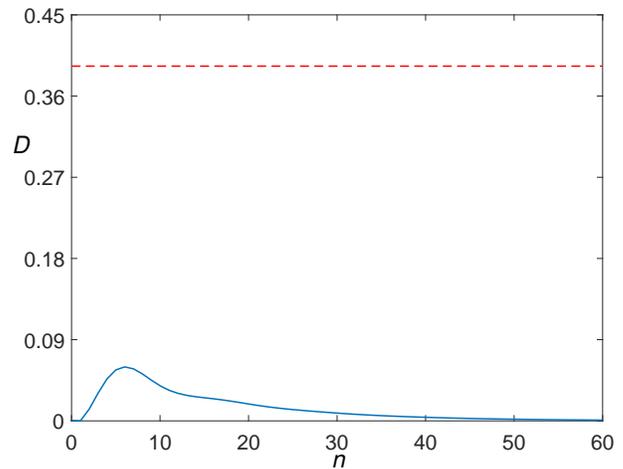}
\caption{(Color online)
Trace distance Eq. (19) (solid blue line) against the number of collisions $n$, a non-Markovian process induced by initial classical correlation between system $S$ and environment ($\xi=0.855$ in Eq. (18)) in the Markovian coupling region of Fig. 2 with $\{\gamma=\frac{\pi}{14}$, $\delta=\frac{\pi}{6}$\}, and the state with coherence of each ancilla of the reservoir ($p=0.4$ in Eq. (16)) which corresponds to a Markovian dynamics in Fig. 3. And the bound of Eq. (20) is indicated by the upper red-dashed line.}
\label{Fig5}
\end{figure}

From the discussion above, it is always able to induce a transition from Markovian to non-Markovian dynamics for initial quantum correlation between system and its environment. For initial classical correlation state (18) and a thermal state of the reservoir, from numerical calculation we find that the trace distance Eq. (19) is always zero with the number of collisions $n$ within the Markovian region of coupling presented in Fig. 2. In order to study the effect of initial classical correlation on NM more comprehensively, we use the measure of the degree of NM in the Appendix (Eq. (A.1)), and we find the similar result that $\mathcal{N}$ in Eq. (A.1) is also zero. However it is worth noting that Eq. (A.1) can only be used to witness the occurrence of non-Markovian dynamics rather than to confirm a Markovian dynamics. Therefore, from now on it cannot guarantee that the dynamics of system must be Markovian for initial state (18). And thus for initial classical correlation we do only claim that the $\mathcal{N}$ in Eq. (A.1) is zero in the case of thermal reservoir comparing to the case of reservoir with coherence (see below). In Fig. 5, we plot the trace distance Eq. (19) against the number of collisions $n$ for initial classical correlation state (18). $\gamma=\frac{\pi}{14}$ and $\delta=\frac{\pi}{6}$, and a state with coherence of each ancilla of the reservoir, $p=0.4$ in Eq. (16), which corresponds to a Markovian dynamics presented in Fig. 3. Obviously the change of trace distance is similar to the case of initial quantum correlation presented in Fig. 4, the trace distance increases from zero to a maximum and then decreases until to zero. This also indicates a non-Markovian dynamics of system and the trace distance here exceeds the initial value. Note that in Ref.\cite{f40} it has pointed out that the effects of initial classical correlation are related to the form of interaction, and they have verified that the existence of the initial classical correlation will not make the trace distance of the system exceed the initial value if two qubits are under the action of controlled-NOT gate only; and if first apply the controlled-NOT gate and then a swap operation, it can obtain a growth of the trace distance. In our case, a growth of the trace distance and a non-Markovian dynamics are emerged by means of coherence of reservoir in the case of initial classical correlation. And Eq. (20) is also satisfied now.

In summary, we study the effect of initial system-environment correlations on system dynamics, including quantum correlation and classical correlation. We realize a growth of the trace distance and a non-Markovian dynamics with the help of initial quantum correlation, however for initial classical correlation this can only be confirmed to occur when there is coherence of the reservoir simultaneously.

\section{Effect of non-Markovianity of systems on thermodynamics}\label{Sec4}
In this section we consider the system is initially in the ground state $|1\rangle$ or excited state $|0\rangle$, and all the environment qubits ($A_{Q}$ and each ancilla of the reservoir) are initially prepared in the same thermal states. In this case the unitary interactions (5) and (7) make the reduced state $\rho^{A_{Q}}_{n}$ maintain the form of thermal state, $\rho^{A_{Q}}_{n}=\rho^{A_{Q}}_{\beta}=\frac{1}{Z}e^{-\beta_{A_{Q}} \hat{H}_{A_{Q}}}$, with the n-dependent inverse temperatures $\beta_{A_{Q}}=\frac{1}{T_{A_{Q}}}$.

It is known that the total von Neumann entropy of the composite system `$S-{A}_{Q}$' under the unitary evolution $U_{S,A_{Q}}$ is invariant during each step, i.e., $S(\rho^{S,A_{Q}}_{n})=S(\tilde{\rho}^{S,A_{Q}}_{n+1})$, here $\tilde{\rho}^{S,A_{Q}}_{n+1}=U_{S,A_{Q}}(\gamma)(\rho^{S,{A}_{Q}}_{n})U^{\dagger}_{S,A_{Q}}(\gamma)$.
Based on this, the change of entropy of system during the $(n+1)$th interaction can be expressed as~\cite{e6}
\begin{equation}
\begin{split}
\Delta\mathrm{S_{n+1}}=&\mathrm{S}(\tilde{\rho}^{S}_{n+1})-\mathrm{S}(\rho^{S}_{n})\\
=&D(\tilde{\rho}^{S,A_{Q}}_{n+1}\parallel\tilde{\rho}^{S}_{n+1}\rho^{A_{Q}}_{\beta})+\textrm{Tr}_{{A}_{Q}}(\tilde{\rho}^{A_{Q}}_{n+1}-\rho^{A_{Q}}_{\beta})\ln\rho^{A_{Q}}_{\beta}\\
&-I(\rho^{S,A_{Q}}_{n}),
\end{split}
\end{equation}
where $\tilde{\rho}^{S}_{n+1}=\mathrm{Tr}_{A_{Q}}[\tilde{\rho}^{S,A_{Q}}_{n+1}]$, $\tilde{\rho}^{A_{Q}}_{n+1}=\mathrm{Tr}_{S}[\tilde{\rho}^{S,A_{Q}}_{n+1}]$, $D(\rho_{1}\|\rho_{2})\equiv \mathrm{Tr}(\rho_{1}ln\rho_{1})-\mathrm{Tr}(\rho_{1}ln\rho_{2})$ is the quantum relative entropy between two density matrices $\rho_{1}$ and $\rho_{2}$, and the mutual information $I(\rho^{S,A_{Q}}_{n})=\mathrm{S}(\rho^{S}_{n})+\mathrm{S}(\rho^{A_{Q}}_{n})-\mathrm{S}(\rho^{S,A_{Q}}_{n})$, measures the correlation between $S$ and $A_{Q}$, and this correlation has been established after
their collision in the first step. According to the definition of $\rho^{A_{Q}}_{\beta}$ above, we can obtain  $\textrm{Tr}_{{A}_{Q}}[(\tilde{\rho}^{A_{Q}}_{n+1}-\rho^{A_{Q}}_{\beta})\ln\rho^{A_{Q}}_{\beta}]=\beta_{{A}_{Q}}\Delta Q_{n+1}$, here
\begin{equation}
\begin{split}
\Delta Q_{n+1}=\textrm{Tr}_{{A}_{Q}}[(\rho^{A_{Q}}_{n}-\tilde{\rho}^{A_{Q}}_{n+1})\hat{H}_{A_{Q}}], \end{split}
\end{equation}
representing the heat flowing from auxiliary qubit ${A}_{Q}$ to system $S$. Therefore, Eq. (21) can also be written as
\begin{equation}
\begin{split}
\Delta\mathrm{S_{n+1}}
=&D(\tilde{\rho}^{S,A_{Q}}_{n+1}\parallel\tilde{\rho}^{S}_{n+1}\rho^{A_{Q}}_{\beta})
+\beta_{{A}_{Q}}\Delta Q_{n+1}-I(\rho^{S,A_{Q}}_{n}).
\end{split}
\end{equation}
Notice that we choose energy-preserving interactions between the bipartite systems, `$S-{A}_{Q}$', `$A_{Q}-\mathcal{R}_{j}$'. Mathematically, this translates as $[\hat{U}_{S,A_{Q}},\hat{H}^{fre}_{S,A_{Q}}]=0$, $[\hat{V}_{A_{Q},\mathcal{R}_{j}},\hat{H}^{fre}_{A_{Q},\mathcal{R}_{j}}]=0$, that is, $[\hat{H}^{int}_{S,{A}_{Q}},\hat{H}^{fre}_{S,{A}_{Q}}]=0$, $[\hat{H}^{int}_{A_{Q},\mathcal{R}_{j}},\hat{H}^{fre}_{A_{Q},\mathcal{R}_{j}}]=0$, here $\hat{H}^{fre}_{S,{A}_{Q}}=\hat{H}_{S}+\hat{H}_{A_{Q}}$, $\hat{H}^{fre}_{A_{Q},\mathcal{R}_{j}}=\hat{H}_{A_{Q}}+\hat{H}_{\mathcal{R}_{j}}$. So that the heat given by the system is completely transferred to the environment, and vice versa. In other words, no heat is given or taken in the form of thermodynamic work while performing the unitary operations. Thus the canonical definition of heat flow $\Delta Q_{n+1}$ in Eq. (22) is valid and compatible with thermodynamics, and the term $\beta_{{A}_{Q}}\Delta Q_{n+1}$ in Eq. (23) is associated with the system entropy change due to heat exchanges.

\begin{figure}[h]
\centering
\includegraphics[scale=0.58]{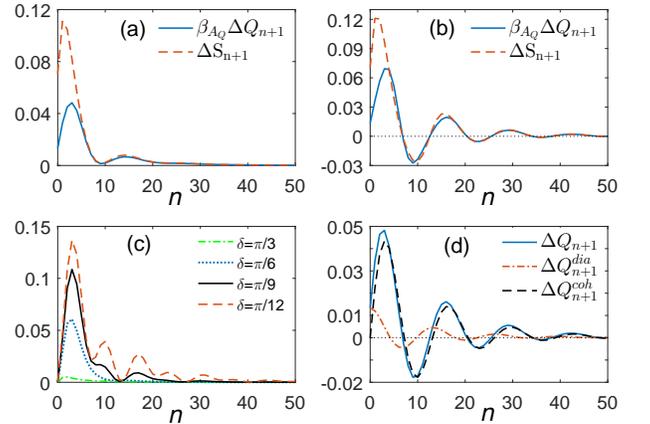}
\caption{(Color online)
(a)-(b): $\Delta S_{n}$ and $\beta_{A_{Q}}\Delta Q_{n+1}$ against the number of collision $n$ with $\gamma=\frac{\pi}{14}$ and different $\delta$, $\delta=\frac{\pi}{6}$ (Markovian region of coupling (a)), $\delta=\frac{\pi}{9}$ (non-Markovian region of coupling (b)). (c) The correlation $I(\rho^{S,A_{Q}}_{n})$ in Eq. (23) against the number of collision $n$ with fixed $\gamma$ ($\gamma=\frac{\pi}{14}$) and different $\delta$. (d) The heat $\Delta Q_{n+1}$ absorbed by system and its two contributions $\Delta Q^{dia}_{n+1}$ and $\Delta Q^{coh}_{n+1}$ against the number of collision $n$, and the parameters are the same as those of (b). For all plots the initial state of system is ground state $|1\rangle$, and the initial states of auxiliary qubit and reservoir qubits are in the same thermal states $\rho_{\beta}$ in Eq. (16) with $T=1$, and $\omega=1$.}
\label{Fig6}
\end{figure}

In order to study the change of entropy of system especially that results from
heat exchanges with different memory effects of environment, in Fig. 6 (a)-(b) we plot $\Delta S_{n+1}$ and $\beta_{A_{Q}}\Delta Q_{n+1}$ against the number of collisions $n$ of a Markovian region of coupling ($\gamma=\frac{\pi}{14},\delta=\frac{\pi}{6}$) in Fig. 6(a) and a non-Markovian region of coupling ($\gamma=\frac{\pi}{14},\delta=\frac{\pi}{9}$) in Fig. 6(b). The initial state of system is ground state $|1\rangle$, and the initial states of auxiliary qubit and reservoir qubits are in the same thermal states $\rho_{\beta}$ in Eq. (16) with $T=1$. It shows that the changes of $\Delta S_{n+1}$ and $\beta_{A_{Q}}\Delta Q_{n+1}$ are almost consistent with the increase of $n$, increasing first and then oscillating decay. However $\Delta S_{n+1}$ and $\beta_{A_{Q}}\Delta Q_{n+1}$ are always larger than zero for Markovian environment (Fig. 6(a)), and which can be less than zero during some time intervals for non-Markovian environment (Fig. 6(b)). Physically this can be understood as following. We define $\rho_{ij}$ ($i,j=1,2,3,4$) are the matrix elements of state $\rho^{S,{A}_{Q}}_{n}$ of `$S-{A}_{Q}$' before their $(n+1)$th collisions. Due to the correlations between $S$ and ${A}_{Q}$, $\Delta Q_{n+1}$ in Eq. (22) can be divided into two different contributions:
\begin{equation}
\Delta Q_{n+1}=\Delta Q^{dia}_{n+1}+\Delta Q^{coh}_{n+1},
\end{equation}
where
\begin{equation}
\begin{split}
&\Delta Q^{dia}_{n+1}=\omega\sin^{2}(\gamma)(\rho_{33}-\rho_{22}),\\
&\Delta Q^{coh}_{n+1}=\omega\mathrm{Im}(\rho_{23}) \sin(2\gamma),
\end{split}
\end{equation}
are the heats determined, respectively, by the diagonal and coherent (off-diagonal) elements of state $\rho^{S,{A}_{Q}}_{n}$, and $\omega=\omega_{S}=\omega_{E}$ is the resonance frequency of $S$, $A_{Q}$ and $\mathcal{R}_{j}$. The nonzero coherent term $\rho_{23}$ of $\rho^{S,A_{Q}}_{n}$ is a direct witness of correlation between $S$ and $\rho^{A_{Q}}_{n}$, which in turn gives the correlation-dependent heat $\Delta Q^{coh}_{n+1}$. For fixed parameter $\gamma$, the relatively large values of $\delta$ lead to Markovian dynamics, and the established system-environment correlations are weak, so the contribution $\Delta Q^{dia}_{n+1}$ plays a major role in determining the behavior of total heat $\Delta Q_{n+1}$. This can be verified by Fig. 6(c): the correlations $I(\rho^{S,A_{Q}}_{n})$ established within the dynamical process decrease with the increase of $\delta$ for fixed $\gamma$. Differently, when $\delta$ is sufficiently small (non-Markovian dynamics) the behavior of $\Delta Q_{n+1}$, especially its transition from positive to negative values, is mainly determined by the contribution $\Delta Q^{coh}_{n+1}$, as showed in Fig. 6 (d).

\begin{figure}[h]
\includegraphics[scale=0.56]{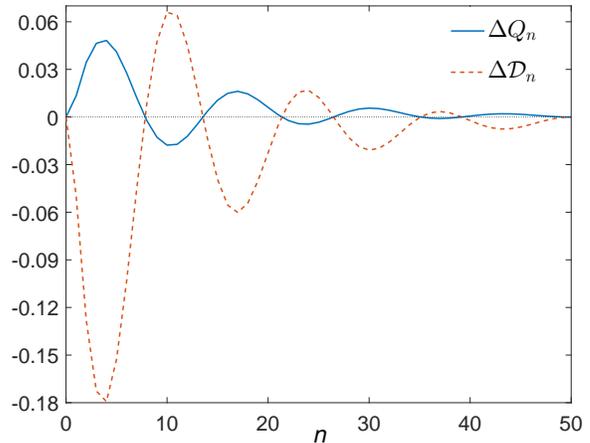}
\centering
\caption{(Color online) Behavior of heat flux $Q_{n}$ and the change of trace distance $\Delta\mathcal{D}_{n}$ against the number of collision $n$ for initial systems states
$\{|0\rangle,|1\rangle\}$, and the other parameters are the same as that given in Fig. 6(b) and Fig. 6(d).}
\label{Fig8}
\end{figure}

As showed above in Fig. 6(d), the direction of heat flux is non-unidirectional if memory effects are present. In Fig. 7 except heat flux we plot the change of trace distance $\Delta\mathcal{D}_{n}$ $(\Delta\mathcal{D}_{n}=\mathcal{D}_{n}-\mathcal{D}_{n-1})$ with $n$ for the initial pair of states $\{|0\rangle,|1\rangle\}$ of system. It shows that the behaviors of heat flux and the change of trace distance are perfectly aligned with one another, i.e., the change of direction of heat flux is commensurate with the onset of a non-Markovian dynamics, and from numerical calculations we find that this is generally true for any choice of initial system states. This feature provides a thermodynamic interpretation of non-Markovianity as quantified by information backflow~\cite{f12} in the case of energy-preserving system-environment interactions. This is because, in such cases, information flow between the system and environment is always accompanied by energy exchange~\cite{c2}.

\section{CONCLUSION}\label{Sec6}
In summary, we have studied a system that is coupled to a structured environment consisting of a auxiliary system and a reservoir. We have showed the possibility of manipulating the non-Markovianity of system of interest, and the system can realize a transition from Markovian to non-Markovian dynamics by different ways, including the coupling strength of system-auxiliary system and auxiliary system-reservoir, initial system-environment correlation and coherence of environment. And we have showed that the trace distance between two states of system increases above its initial value in the presence of initial system-environment correlations. Especially the growth of trace distance is emerged by means of coherence of reservoir in the case of initial classical correlation, and this is different from the result showed in Ref.\cite{f40} that the effects of initial classical correlation are related to the form of interaction.

By studying the entropy change of system especially that results from heat exchanges with different memory effects of environment, we have revealed that the essence of entropy change between positive and negative values during non-Markovian evolution is due to the contribution of heat flux determined by coherence. Then we have found that the dynamic behaviors of heat flux and the change of trace distance are perfectly aligned with each other, in other words information flow between the system and environment is always accompanied by energy exchange.

Note that in this paper we have used the collision model to investigate the influences of non-Markovian dynamics, and the relation of non-Markovianity and thermodynamics. The reason to consider this simple model is that exact solutions can be obtained for a general class of initial system-environment correlations and the initial states of reservoirs with coherence. We expect that some features of the non-Markovianity and thermodynamics in this simple model might be similar to those in more involved but less tractable structured-environment models, so we can gain some insight into the general feature of the effects of initial system-environment correlations and reservoirs with coherence on non-Markovianity, and hence the relation between non-Markovianity and thermodynamics.

\begin{acknowledgments}
We thank Jian Zou, Zhong-Xiao Man and Chao-Quan Wang for discussions. This work is supported by The National Natural Science Foundation of China (Grant No. 11947047).
\end{acknowledgments}

\appendix*
\renewcommand{\appendixname}{APPENDIX~}
\section{Non-Markovianity witness with initial classical correlation}
We introduce the degree of NM  which makes use of the non-monotonicity of the trace distance between two states of system to witness the effect of initial classical correlation on the non-Markovian dynamics of the system~\cite{ll10},
\begin{equation}
\begin{split}
\mathcal{N}=\mathop{max}\limits_{\theta\in[0,\pi],\varphi\in[0,2\pi]}\sum_{n\in\sigma^{+}}\Delta\mathcal{D}(n+1),
\end{split}
\end{equation}
where $\Delta\mathcal{D}(n+1)=\mathcal{D}[\rho_{s}(n+1),\tilde{\rho}_{s}(n+1)]-\mathcal{D}[\rho_{s}(n),\tilde{\rho}_{s}(n)]$, and $\rho_{s}(n+1)$ is the same as that in Eq. (19), $\tilde{\rho}_{s}(n+1)$ is the reduced state of system after the $(n+1)$th interaction corresponding to the initial state $\rho_{S,A_{Q}}(0)=\tilde{\rho}_{s}(0)\otimes\mathrm{Tr}_{S}(\rho_{S,A_{Q}}^{2}(0))$ with the initial system state
$\tilde{\rho}_{s}(0)=\cos\frac{\theta}{2}|0\rangle+e^{i\varphi}\sin\frac{\theta}{2}|1\rangle, \theta\in[0,\pi],\varphi\in[0,2\pi]$. The maximization is performed by taking all possible system states $\tilde{\rho}_{s}(0)$ over the Bloch sphere. And here the definition of $\sigma^{+}$ is the same as that in Eq. (13), in which $\Delta\mathcal{D}(n)>0$.

\end{document}